\documentclass[pre,twocolumn,floatfix]{revtex4}
\usepackage{graphicx}

\begin{document}
\title{Elimination of cracks in self-assembled photonic band gap crystals}

\author{A.A.~Chabanov}
\author{Y.~Jun}
\author{D.J.~Norris}
\email[electronic address: ]{dnorris@umn.edu}
\affiliation{Department of Chemical Engineering and Materials Science,
University of Minnesota, Minneapolis, MN 55455}

\date{December 15, 2003}

\begin{abstract}
Thin colloidal crystals (or synthetic opals) composed of St\"{o}ber silica spheres typically develop cracks when they are utilized to obtain photonic band gap crystals (or inverted opals). We find that by sintering the silica spheres prior to assembly of the opal these cracks can be avoided. We report the effects of temperature and duration of the heat treatment on 850 nm silica spheres using electron microscopy, thermogravimetry, and light scattering. We also find a large dependence of the refractive index of the silica on the temperature of the heat treatment. This may allow tuning of the refractive index of silica spheres.
\end{abstract}

\maketitle 
Photonic band gap crystals are synthetic materials in which a certain range of electromagnetic frequencies is not allowed to exist \cite{Yablon87,John87}. They have been studied because this property, known as the photonic band gap, can provide a versatile means to control optical signals \cite{Joanop97,book}. However, to obtain a photonic band gap for optical frequencies, one must be able to fabricate structures that are three-dimensionally periodic on an optical length scale and composed of a solid with a high refractive index ($n$). One approach to achieve this is through the self-assembly of spherical colloidal particles \cite{Vlasov95,Pine97,Velev97,Stein98,Vos98,Zahidov98,Braun02}. Micrometer-scale silica spheres that are nearly monodisperse can be induced to organize into crystals known as synthetic opals. The spheres are close-packed on a face-centered cubic (fcc) lattice. To obtain a photonic band gap, these opals must be infiltrated with a material with a large $n$ and the spheres must then be selectively removed by a wet etchant. The resulting structure, referred to as an inverted opal, has been predicted to have a photonic band gap if the infill has $n>2.85$ \cite{Sozuer92,Busch98}. Consequently, chemical vapor deposition (CVD) has been used to obtain silicon ($n$=3.5) inverted opals to explore the photonic band gap in self-assembled materials \cite{Blanco00,Norris01}.
  
One issue of concern is that, as with any crystal growth process, a variety of defects (e.g., vacancies, stacking faults, interstitials, etc.) can form during the self-assembly of synthetic opals. These are transferred from the opal to the photonic band gap crystal and, depending on the type of defect, may render it unsuitable for applications \cite{Zhang00}. Thus, one focus of recent work has been to reduce defect densities in these structures \cite{vanBlaad97}. In particular, methods have been recently developed to deposit thin opals ($\sim$20 layers) with low defect densities on a flat substrate \cite{Colvin99}. Capillary forces in the meniscus between the substrate and a colloidal sol induce the crystallization \cite{Denkov93}. Using these methods, thin silicon inverted opals with optical properties consistent with a photonic band gap have recently been grown \cite{Norris01}.

However, less attention has been focused on eliminating defects that are formed during the infiltration step. In particular, cracks can appear in thin opaline films under conditions which induce shrinkage of the spheres. For example, in opals assembled from polymer spheres, this can occur due to electron-beam exposure in an electron microscope \cite{Colvin99}. Opals assembled from silica spheres (as discussed here) are more resistant to shrinkage, but shrinkage still occurs if the opal is heated to elevated temperatures. Thus, when opals are infiltrated with silicon at temperatures above 500$^0$C \cite{Blanco00,Norris01}, cracks appear which are transferred to the inverted opals. For applications in photonics, these cracks are a serious problem that can limit their use.
 
Here, we explore methods to eliminate these cracks. Namely, we show that by pre-shrinking the silica spheres prior to assembly of the opal, the spheres do not shrink significantly during the silicon infiltration step. Thus, inverted opals can be obtained in which the cracks have been eliminated. Two alternative methods to achieve pre-shrinkage of the silica spheres have been studied. First, the spheres were refluxed for prolonged periods in high-boiling-point solvents. Second, the spheres were sintered as dry powders and then redispersed. Below we report the details of our findings.

Our silica particles were produced by the standard St\"{o}ber synthesis \cite{Stober} which was modified to obtain particles $>$600 nm in diameter \cite{Giesche94}. As determined by scanning electron microscopy (SEM), the particles used here had an average diameter of 850 nm with a standard deviation of 1.4\%. They were obtained from the same batch discussed in Ref.~\cite{Norris01}. After growth, the particles were redispersed 3 times in deionized water and stored in absolute ethanol. To assemble an opal, the particles were resuspended in methanol ($\sim$1\% by volume) in a small vial, which was then placed in a sand bath. A temperature gradient was applied across the vial (50$^0$C at the bottom to 40$^0$C at the top) and a substrate was submersed in the suspension. As the solvent evaporated, a thin opal was deposited and occupied a centimeter-scale area on the substrate. The opal was then dried and studied with SEM. The opals had a structure similar to previous reports \cite{Norris01,Colvin99}. In particular, the opals had no discernible cracks after growth.

Silicon inverted opals were obtained as reported previously \cite{Norris01}. In brief, the opal was infiltrated with Si using low-pressure CVD (LPCVD). During the process the sample temperature was maintained at 550$^0$C for 5 hours. After infiltration, the top surface of the opal was exposed by reactive ion etching and the silica spheres were removed with a 5\% aqueous solution of HF. A SEM image of a typical Si inverted opal obtained by this process is shown in Fig.~1a. The image indicates that the opal has developed numerous cracks during the deposition process. Moreover, since the cracks are filled with Si, this indicates that the silica spheres shrank and formed cracks early in the process. This implies that if the opal were assembled of pre-shrunken silica spheres perhaps no cracks would develop, thereby suggesting a possible solution to the crack problem.

Our first attempt to pre-shrink the spheres involved refluxing the spheres in high-boiling-point solvents. For example, by dispersing the silica in ethylene glycol, the spheres could be heated to $\sim$260$^0$C for prolonged periods. However, we found that at elevated temperatures the pH of the dispersion quickly rose, presumably due to release of adsorbed hydroxide ions. At high pH, the size of the spheres was reduced. However, while our goal is to densify the spheres, they were actually just etched due to the higher solubility of silica under these highly basic conditions. Thus, such spheres would still shrink during the high temperature infiltration step. Attempts to eliminate this etching with repeated washings of the initial dispersion or addition of buffers were not successful.

Instead, we moved to another approach --- sintering the spheres as dry powders and then redispersing them. This approach was successful, as shown below, and a systematic study of the variation in size and density of the spheres due to variations in time and temperature of the heat treatment were performed to understand the dynamics of the process. To determine changes in mass, loose powders were heated at constant temperature in static air for up to 10 hours using a Perkin-Elmer TG7 thermogravimeter. Changes in the size of the spheres were measured with SEM. In Fig.~2a we plot the relative volume loss of the spheres with time. The reported time does not include that initially required to ramp the sample up to the final temperature (at 20$^0$C per min). Thus, values of volume loss are finite at zero time. The data show that after a rapid initial loss, the volume of the silica spheres approaches a constant value. However, the net volume loss increases with increasing temperature, $T$. Also, the time at which the volume levels off, decreases with increasing $T$. At 600$^0$C, the total volume loss of silica spheres of $\sim$20\% is achieved in 4 hours. In Fig.~2b, we plot the relative weight loss of the spheres with time. The weight loss is seen to be correlated with volume changes. At 600$^0$C, the total weight loss of silica spheres is $\sim$12\%, which is in agreement with previous thermogravimetric studies of silica particles \cite{Sacks84,Badley90}. As these results show that the weight loss is less than the volume loss, they indicate that the silica spheres are becoming denser. This was confirmed using a helium pycnometer to measure the density of our silica powders. Whereas the density of the original St\"{o}ber silica spheres was 1.97$\pm$0.05 g/cm$^3$, it increased to 2.10$\pm$0.05 g/cm$^3$ and 2.17$\pm$0.05 g/cm$^3$ with heat treatments at 300$^0$C and 600$^0$C, respectively, in agreement with previous measurements \cite{Giesche94,Sacks84,Badley90}. The increase in the density of silica spheres is attributed to the removal of water and ethanol from the silica during heating and elimination of pores (see more below) \cite{Iler}.
\begin{figure}
\includegraphics [width=3in] {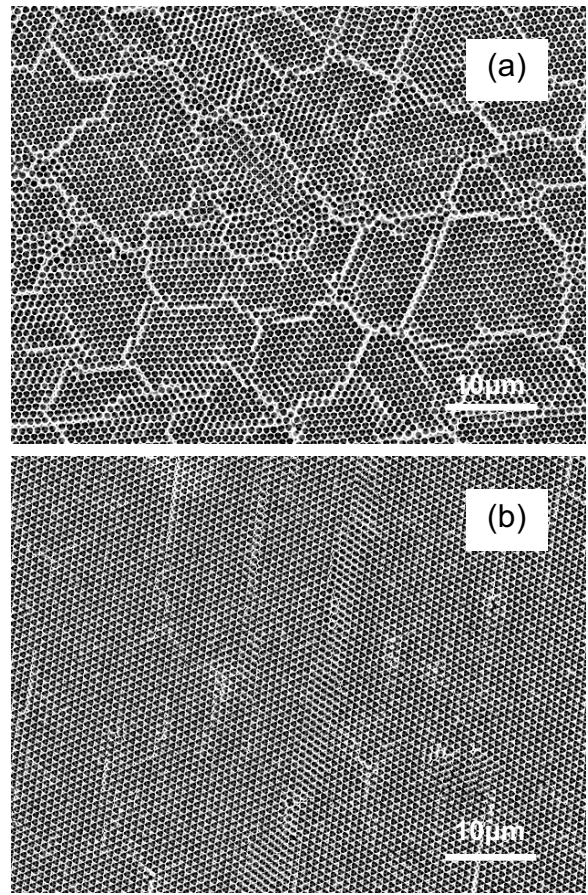}
\caption{(a) SEM image of a heavily-cracked Si inverted opal. The opal template for the crystal was assembled of 850 nm silica spheres; the cracks appeared during Si LPCVD due to shrinkage of the spheres. (b) SEM image of a crack-free Si inverted opal. The spheres were sintered as a dry powder at 600$^0$C for 4 hours prior to assembly of the opal; after the heat treatment, the diameter of the spheres was 792 nm.}
\end{figure}
 
Based on the results of Fig.~2, we prepared silica spheres for opal assembly by heating silica powder at 600$^0$C for 4 hours. It is important to note that SEM measurements indicated that the heat treatment caused no significant change in the polydispersity of the silica spheres. The particles shrunk from 850 nm $\pm$ 1.4\% to 793 nm $\pm$ 1.4\%. After the sintering, the particles were redispersed in methanol using ultrasonication. Silica aggregates left after sonication were removed by sedimentation. Then, Si inverted opals were fabricated in the same manner as described above. A SEM image of the resulting inverted opal is shown in Fig.~1b. No cracks have developed in the inverted opal formed from the pre-shrunken spheres.  Thus, our approach provides a solution to the crack problem in self-assembled photonic band gap crystals.
\begin{figure}
\includegraphics [width=\columnwidth] {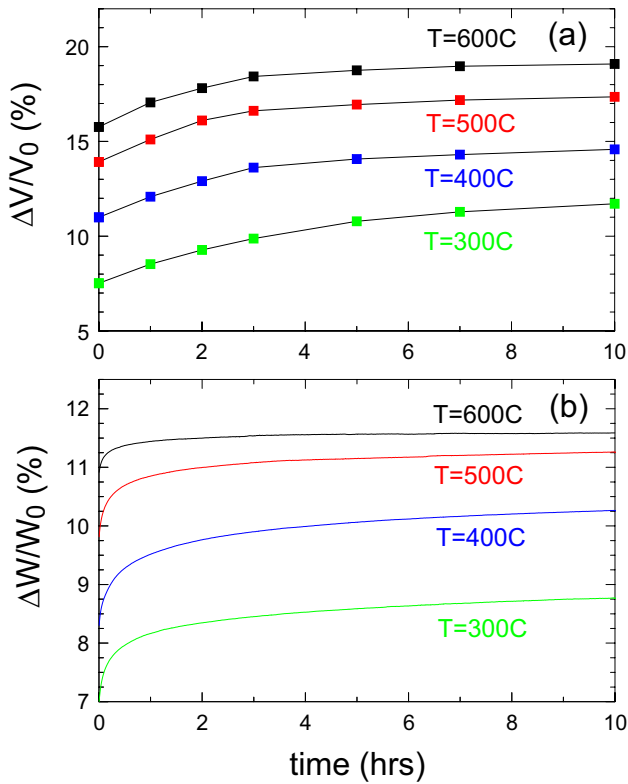}
\caption{(a) Relative volume loss and (b) relative weight loss of silica spheres heated at $T$=300, 400, 500, and 600$^0$C plotted versus the time of the heat treatment.}
\end{figure}

Finally, we note that, during our study, we also discovered some unexpected behavior in the refractive index of the silica particles. To obtain additional evidence of the densification of the silica after heat treatment, we determined the refractive index of the silica spheres, $n_{sp}$, by measuring the optical transmittance through dilute dispersions in mixtures of ethanol ($n_{e}$=1.360 at 589 nm) and 1-bromonaphthalene ($n_{b}$=1.657 at 589 nm). In this case, which is in the single scattering limit, the attenuation of the transmittance due to scattering is a function of the refractive index contrast between the silica spheres and the solvent. The transmittance is at a maximum when the refractive index of the solvent matches $n_{sp}$. Using a UV-visible spectrometer, transmittance spectra were obtained for 7 different values of the refractive index contrast, slightly above and below the value of unity, by changing the concentration of 1-bromonaphthalene in ethanol. After correcting for attenuation due to absorption by the solvents, transmittance at 589 nm was fit by a polynomial of the refractive index of the solvent. This polynomial was then solved for the maximum transmittance, corresponding to a refractive index contrast of one.
  
The results of the analysis are presented in Fig.~3, where $n_{sp}$ is plotted versus $T$. According to our density measurements of the original and heat-treated silica powders, one might expect an increase in the refractive index of silica spheres with increasing $T$. However, $n_{sp}$ of the heat-treated silica is actually smaller than that of the original silica and shows an unusual dependence upon $T$.  The refractive index first decreases with increasing $T$; then, at $T>400^0$C, it increases with $T$, approaching the refractive index of fused quartz, indicated by the dashed line.  This unusual behavior may be explained by the ultramicroporous structure of silica \cite{Iler,vanHelden81}. During the heat treatment, we propose that two processes are occurring: material is being eliminated from the pores (which decreases the effective index of the particles) and the pores are eliminated entirely through condensation (which increases the effective index of the particles). A competition between these two effects can explain the results in Fig.~3.
\begin{figure}
\includegraphics [width=\columnwidth] {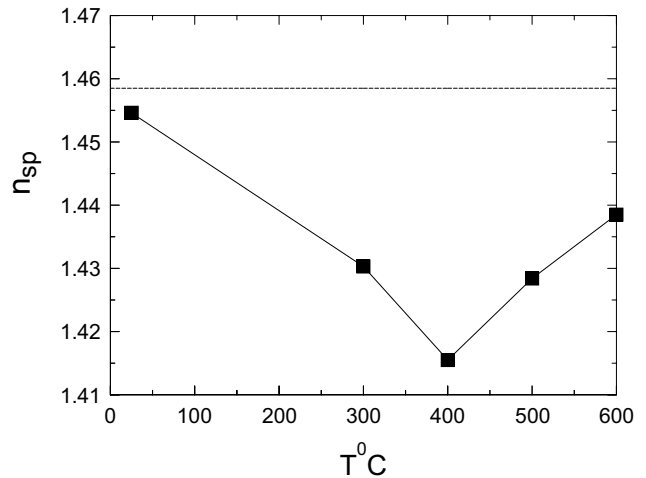}
\caption{Refractive index at 589 nm of the original silica spheres (as synthesized) and silica spheres heat-treated at $T$=300, 400, 500, and 600$^0$C for 10 hours. The dashed line indicates the refractive index of fused quartz.}
\end{figure}

Even more surprising, we found that $n_{sp}$ depended upon the storage history of the silica spheres. For example, silica spheres stored in water and then sintered at 600$^0$C for 10 hours exhibited $n_{sp}$ of 1.36. This is quite different from $n_{sp}$ of 1.44 obtained under the same sintering conditions for spheres stored in ethanol (Fig.~3). These results were very reproducible and may provide one reason why reported values of $n_{sp}$ can vary significantly from study to study.  However, future work is necessary to completely clarify these effects.

In conclusion, we report a simple method to avoid cracks in silicon-inverted opals. By sintering silica spheres as dry powders at temperatures as high as 600$^0$C prior to the assembly of the opal, cracks do not appear subsequently during the silicon infiltration step. Thus, this method eliminates another class of important defects in these materials. In addition, we find that the heat treatment of the silica spheres can significantly change their refractive index, depending on the temperature and their solvent history. With further understanding, this effect may be useful for controlling the refractive index of silica particles, potentially over a range between 1.36 and 1.44.

We would like to thank Jing Wang for assistance in measuring refractive index of silica spheres. Y.J. acknowledges a partial research fellowship from the Industrial Partnership for Research in Interfacial and Materials Engineering (IPRIME) at the University of Minnesota. This work was supported in part by the MRSEC Program of the National Science Foundation under DMR-0212302.

\end{document}